\documentclass[aps,prl,superscriptaddress,twocolumn,a4,showpacs]{revtex4}
\usepackage{amsmath, amssymb, psfrag, tabls, dcolumn, bm, color}

\usepackage{graphicx}  
\usepackage{dcolumn}   
\usepackage{amsopn}

\newcounter{subeqn} %

\begin{document}
\title{Bosonic transport through a chain of quantum dots}

\author{Anton Ivanov }
\address{Institut f\"ur Theoretische Physik, Universit\"at  Heidelberg, Philosophenweg 19, 69120 Heidelberg, Germany}
\author{Georgios Kordas}
\address{Institut f\"ur Theoretische Physik, Universit\"at  Heidelberg, Philosophenweg 19, 69120 Heidelberg, Germany}
\author{Andreas Komnik}
\address{Institut f\"ur Theoretische Physik, Universit\"at  Heidelberg, Philosophenweg 19, 69120 Heidelberg, Germany}
\author{Sandro Wimberger}
\address{Institut f\"ur Theoretische Physik, Universit\"at  Heidelberg, Philosophenweg 19, 69120 Heidelberg, Germany}

\begin{abstract}
The particle transport through a chain of quantum dots coupled to two bosonic reservoirs is studied.
For the case of reservoirs of non-interacting bosonic particles, we derive an exact set of stochastic differential equations, whose memory kernels and driving noise are characterised entirely by the properties of the reservoirs. Going to the Markovian limit an analytically solvable case is presented. The effect of interparticle interactions on the transient behaviour of the system, when both reservoirs are instantaneously coupled to an empty chain of quantum dots, is approximated by a semiclassical method, known as the Truncated Wigner approximation. The steady-state particle flow through the chain and the mean particle occupations are explained via the spectral properties of the interacting system.
\end{abstract}
\date{April 19, 2013}

\maketitle

\section{1 Introduction}
\label{intro}

The advance in technology in the last decades has allowed the creation of increasingly smaller devices reaching the point where the realisation of 
logic structures on the atomic level is possible \cite{Smith95,Terabe05,Schimmel04,Fuechsle}.
Because of their low dimensionality and temperature the dynamics of the system can be dominated by quantum effects, opening a large playground for experimental testing of many-body correlation effects on particle (charge or mass) transport. 

These ideas boosted also the investigation of transport of ultracold atoms in systems with reduced dimensionality. 
Transport of fermionic and bosonic ultracold atoms in quantum wires and in one-dimensional optical lattices is studied theoretically in \cite{Ax,Schlagheck10,Schlagheck13,Chien12,Bruderer,Chien13}. In \cite{A31} a possible realisation of an atom analogue of an electron quantum point contact by the use of a microfabricated magnetic wave\-guide is presented. In an experiment a macroscopic  atomic cloud was divided into two reservoirs separated by a narrow channel by the use of a laser beam \cite{Esslinger12}, thus creating a cold-atom analog of a mesoscopic conductor. Recent advances in the manipulation of cold atoms loaded in optical lattices are presented in \cite{Schlosser12}. Decreasing the dimensionality of the tunneling to zero, a new field is investigated - the atomtronics. The creation of bosonic analogues to the mesoscopic 
systems used in electronic devices like a diode or field-effect transistor is suggested in \cite{Anderson12} and also theoretically investigated in \cite{Pepino10,Gajdacz12}.

In this work we focus on bosonic transport through a chain of quantum dots coupled to two bosonic reservoirs that keep the system far from equilibrium. Given the by now very well understood behaviour of electronic (fermionic) systems, the first obvious question is about the differences between the bosonic and fermionic transport. It is known that the fermionic Anderson impurity model -- a quantum dot with few energy levels, coupled to two electrodes (electron baths) is the simplest possible model for a field effect transistor (FET). Since the ultracold gas based systems offer a much better degree of `designability' and coherence control, it is also natural to investigate the possibility of a bosonic FET. Having these goals in mind we offer a formal framework for investigation of such systems on the one hand, and on the other hand propose a number of efficient and physically meaningful approximation techniques, which are able to treat even interacting systems. 

In Section 2 we derive a set of stochastic differential equations for the time evolution of the reduced system by writing down the Keldysh partition function of the system and integrating out the reservoir degrees of freedom. In order to derive the set of equations one performs essentially the same steps as in \cite{polk03a}, where a closed system is considered, the difference being only in the addition of two bosonic reservoirs. 
In Section 3 we restrict the system to the special case of two bosonic Markovian reservoirs, which is analytically solvable in the noninteracting case. 
In Section 3.1 we focus on the steady state properties of the system. New effects, appearing after an addition of an interparticle interaction term to the system Hamiltonian, are explained by the use of the spectral properties of the chain of quantum dots. A possible solution in the strongly interacting limit is also suggested. In Section 3.2 the transient behaviour of an initially empty chain of quantum dots, which is instantaneously coupled to two Markovian reservoirs, is calculated. We find a simple scaling law between the time needed to reach a steady state and the strength of the inteparticle interaction. Section 4 concludes the paper, offers a possible experimental realisation of our setup, 
and outlines the avenues for further research.

\section{2 General derivation of a stochastic differential equation}
\label{sec:1}

\subsection{2.1 Single quantum dot coupled to a bosonic reservoir}

To start with we consider a system consisting of a single quantum dot at energy $\Delta$ coupled to a bosonic reservoir with spectral density $\mathcal{D}(\omega)$ and occupation of the modes $n(\omega)$. At the initial time $t_i$ the density matrix of the system is assumed to be a direct product of the density matrices of the reservoir $\hat{\rho}$ and the quantum dot $\hat{\sigma} $. The reservoir is modeled as a set of noninteracting harmonic oscillator levels. 
Their eigenfrequencies $\varepsilon_k$ should form a continuum, which ensures that the time evolution is irreversible and a steady state is reached. The Hamiltonian of the system is given by
\begin{equation}
\begin{array}{rcl}
\hat{H} & = & \Delta \hat{a}^{\dagger}_{}\hat{a}  - \sum \limits^{}_{k} \gamma^{}_{k}\big( \hat{a}^{\dagger}_{}\hat{L}^{}_{k} + \hat{L}^{\dagger}_{k}\hat{a}  \big) + \sum \limits^{ }_{k} \varepsilon^{}_{k}\hat{L}^{\dagger}_{k}\hat{L}^{}_{k},
\end{array}
\end{equation}
where $\hat{a}^{\dagger}_{},\hat{L}^{\dagger}_k $ create a particle in the quantum dot or in the reservoir mode $ k $. One can write down the Keldysh partition function \cite{Kamenev07}, which in the continuous time notation is given by
\begin{equation}
\begin{array}{rcl}
\mathcal{Z} & = & \int \prod \limits^{ }_{k } D[\textbf{L}^*,\textbf{L}] \langle L_{k,-}(t_i) | \hat{\rho}_{k} | L_{k,+}(t_i) \rangle  \\
 & & \times \int D[\textbf{a}^*, \textbf{a}] \langle a_{-}(t_i) | \hat{\sigma} | a_{+}(t_i) \rangle \\[1.0mm]
 &  & \times e^{-L^{*}_{k,-}(t_i) \cdot L^{}_{k,-}(t_i) }   e^{-a^{*}_{-}(t_i) \cdot a^{}_{-}(t_i) }  e^{i\mathcal{S}} ,\\
 \textbf{L}_{k}(t) & = & ( L^{}_{k,-}(t) , L^{}_{k,+}(t) )^T_{}, \\
 \textbf{a}(t) & = & ( a^{}_{-}(t) , a^{}_{+}(t) )^T_{} .
 \end{array} 
\end{equation}
The $-/+$ subscript denotes the position of the field on the forward/backward branch of the Keldysh contour and the ket-vectors $ | a \rangle, | L_{k} \rangle $ are 
eigenvectors of the annihilation operators $\hat{a}$ and $\hat{L}_{k}$.  The initial time on both branches of the Keldysh contour is denoted by $t_i$ and its turning point by $t_f$. The corresponding action is given by
\begin{equation}
\begin{array}{rcl}
\mathcal{S} & = & \int^{t_f}_{t_i} d\tau \big \lbrace \textbf{a}^{\dagger}(\tau) g^{-1}(\tau) \textbf{a}(\tau) +
				  \sum \limits^{ }_{k }  \textbf{L}^{\dagger}_{k}(\tau) g^{-1}_{k}(\tau) \textbf{L}^{}_{k}(\tau) \\
 & &
				 + \sum \limits^{ }_{k }  \gamma^{}_{k} \big( \textbf{L}^{\dagger}_{k}(\tau) \sigma^{}_z  \textbf{a}(\tau)    + \textbf{a}^{\dagger}_{}(\tau) \sigma^{}_z \textbf{L}^{}_{k}(\tau) \big) \big\rbrace .
\end{array}
\end{equation}
where $g^{-1}(\tau) = (i\partial^{}_{\tau} - \Delta )\sigma^{}_z $, $g^{-1}_k (\tau) = (i \partial^{}_{\tau} - \varepsilon^{}_k)\sigma^{}_z $ and $\sigma^{}_z$ is the Pauli $z$-matrix. 

If one uses the discrete time notation, one can include $ \langle L_{k,-}(t_i) | \hat{\rho}_k | L_{k,+}(t_i) \rangle$ $ e^{-L^{*}_{k,-}(t_i) \cdot L_{k,-}(t_i) }  $ into 
the time discrete form of the matrix $g^{-1}_k(\tau)$  \cite{Kamenev07} and integrate out the reservoir degrees of freedom, thus giving the final result
\begin{equation}
\begin{array}{rcl}
\mathcal{Z} & \! = \! & \int D[\textbf{a}^*,\textbf{a}] e^{-a^*_{-}(t_i)a^{}_{-}(t_i) } 
				\langle a_{-}(t_i) | \hat{\sigma} | a_{+}(t_i) \rangle e^{i \mathcal{S}'} , \\[2.0mm]
\mathcal{S}' & \! = \! &  \int^{t_f}_{t_i}d\tau_1 d\tau_2 \textbf{a}^{\dagger}(\tau_1)  G^{-1}_{}( \tau_1, \tau_2 )  \textbf{a}(\tau_2) ,\\
G^{-1}_{}( \tau_1, \tau_2 ) & \! = \! & \delta(\tau_1 - \tau_2)g^{-1}(\tau_1) -  \sum \limits^{ }_{k } \gamma^2_{k}\sigma^{}_z g^{}_k(\tau_1 - \tau_2)\sigma^{}_z .

\end{array}
\end{equation}
The expectation value of a normally ordered observable $\hat{\mathcal{O}} \equiv \mathcal{O}(\hat{a}^{\dagger},\hat{a})$  at the turning point $t_f$ of the contour is given by
\begin{equation}
\label{eq:Exp_val_Obs}
\begin{array}{rcl}
\langle \hat{\mathcal{O}}(t_f) \rangle & = & \int D[\textbf{a}^*, \textbf{a}] \big\lbrace
 \langle a_{-}(t_i) | \hat{\sigma} | a_{+}(t_i) \rangle
e^{-a^{*}_{-}(t_i) \cdot a^{}_{-}(t_i) } \\[1.5mm]
 & & \times
  \mathcal{O}(a^*_{+}(t_f),a^{}_{-}(t_f))
e^{i\mathcal{S}'} \big\rbrace.
\end{array}
\end{equation}
In the same way as in \cite{polk03a}, where the case of a closed system is considered, one can apply the Wigner transformation
 ($a_{\mp}(\tau) = \psi(\tau) \pm \frac{1}{2} \eta(\tau)$) and integrate out the $\eta(t_i), \eta(t_f)$ fields to reduce Eq.(\ref{eq:Exp_val_Obs}) to
\begin{equation}
\begin{array}{rcl}
\langle  \hat{\mathcal{O}}(t_f) \rangle & = & \int D[\psi^*, \psi, \eta^*, \eta] \big\lbrace \sigma_{\mathcal{W}}(\psi^*(t_i) ,\psi (t_i) ) \\[1.5mm]
 &   &  \times   \mathcal{O}_{\mathcal{W}}(\psi^*(t_f),\psi^{}(t_f)) e^{i\mathcal{S}''} \big\rbrace.
\end{array}
\end{equation}
The Wigner transform of the density matrix  $ \sigma_{\mathcal{W}}(\psi^*(t_i) ,\psi (t_i) ) $ and the Weyl 
symbol of the observable $\mathcal{O}_{\mathcal{W}}(\psi^*(t_f),\psi^{}(t_f))$ are both obtained after integrating out the $\eta^*(t_i),\eta(t_i)$ and $\eta^*(t_f),\eta(t_f)$-fields respectively
\begin{equation}
\begin{array}{rcl}
\sigma_{ \mathcal{W} }(\psi^*, \psi) &  = & \int \frac{d\eta^* d\eta}{4\pi^2} \hspace{1.0mm} \big\lbrace \langle \psi \! + \! \eta/2 |\hat{\sigma} | \psi \! - \! \eta/2 \rangle \\[1.5mm]
 & & \times  e^{-|\psi|^2 - 1/4|\eta|^2 +1/2(\eta^* \psi - \eta \psi^*) } \big\rbrace , \\[1.5mm]
\mathcal{O}_{\mathcal{W}}(\psi^*, \psi) & = &\int \frac{d\eta^*d\eta}{2 \pi} \hspace{1.0mm} e^{-|\eta|^2/2} {\mathcal{O}}( \psi^* \! - \! \eta^*/2, \psi \! + \! \eta/2).
\end{array}
\end{equation}
Calculating $\mathcal{O}_{\mathcal{W}}$ is equivalent to writing down the normal ordered operator in a symmetrised form and then replacing $\hat{a}^{\dagger},\hat{a}$ with 
$\psi^*,\psi$, respectively. The new action has the form:
\begin{equation}
\begin{array}{l}
\mathcal{S}'' =   i \int^{t_f}_{t_i} d\tau_1 d\tau_2 \big\lbrace \eta^*(\tau_1) 2i \big(\Gamma n+\Gamma/2 \big) (\tau_1 - \tau_2)  \eta(\tau_2)   \\[1.9mm]

+ \psi^*(\tau_1)[ \delta( \! \tau_1 \! - \! \tau_2 \! )(i\partial_{\tau_2} - \! \Delta)    - 2i\Gamma( \! \tau_1 \! - \! \tau_2 \! )\Theta(  \! \tau_2 \! - \! \tau_1 \!  )  ]\eta(\tau_2)    \\[1.9mm]

+ \eta^*(\tau_1)[ \delta( \! \tau_1 \! - \! \tau_2 \! )(i\partial_{\tau_2} - \! \Delta)    + 2i\Gamma( \! \tau_1 \! - \! \tau_2 \! )\Theta(  \! \tau_1 \! - \! \tau_2 \!  )  ]\psi(\tau_2) \big\rbrace .
\end{array}
\end{equation}
where $\Gamma(t) = \pi \int^{\infty}_{-\infty} \frac{d\omega}{2 \pi} \mathcal{D}(\omega) \gamma^2(\omega) e^{-i\omega t}  $ 
and $ \big( \Gamma n \big) (t) = \pi \int^{\infty}_{-\infty} \frac{d\omega}{2 \pi} \mathcal{D}(\omega) \gamma^2(\omega) n(\omega) e^{-i\omega t}  $. 
In the noninteracting case the action contains only terms that are linear or quadratic in the $\eta,\eta^*$ fields. 
Both types of terms can be integrated out to give the following result:
 \begin{equation}
\begin{array}{rcl}
\langle  \hat{\mathcal{O}}(t_f)  \rangle & = & \hspace{3.0mm} \int D[\xi^*, \xi ]
e^{-  \sum_{lk} \xi^{*}_l  \Sigma^{-1}_{lk}  \xi^{}_{k} }\\[1.5mm]
   &    & \times  \int D[\psi^*, \psi] \big\lbrace \sigma_{\mathcal{W}}(\psi^*(t_i), \psi(t_i) ) \\[1.5mm]
 & & \times \mathcal{O}_{\mathcal{W}}(\psi^*(t_f), \psi(t_f)  ) \delta( f_1(\psi) )\delta( f_2(\psi^*) ) \big\rbrace, \\[1.5mm]
\Sigma^{}_{lk} & = & 2 (\Gamma n + \Gamma/2)(t_l-t_k). 
\end{array}
\end{equation}
In order to derive the last expression we have divided the time interval into $N$ equal parts $\Delta t = \frac{t_f-t_i}{N}$ ($t_l = t_i + l\cdot \Delta t$) and defined $\Sigma \in \mathbb{C}^{N+1 \hspace{0.1mm} \times \hspace{0.1mm} N+1}$ ($ \Sigma_{lk} \equiv \Sigma(t_l - t_k) $), $  \vec{ \xi } , \vec{ \xi }^* \in \mathbb{C}^{N+1}$ ($\xi_l \equiv \xi(t_l)$).
The time evolution of $\psi$ is determined entirely from the argument of the $\delta$-function. 
If one sets $f_1(\psi)$ to be equal to zero one obtains the following stochastic differential equation:
\begin{equation}
\label{eq:Langevin_eq}
\begin{array}{c}
\partial_{t}\psi(t ) = - i \Delta \psi( t ) - \int^{t}_{t_i} d\tau  2\Gamma( \! t \! - \! \tau \! ) \psi(\tau ) + \zeta( t ),
\end{array}
\end{equation}
where $\zeta(t)$ is a Gaussian stochastic process with zero mean and autocorrelation function given by 
$ \langle \zeta(t) \zeta^{\dagger}(t') \rangle = \Sigma( t - t' )$. The equation of motion for $\psi^*(t)$ is obtained by setting $f_2(\psi^*)$ equal to zero and it is equal to the complex conjugate of Eq. (\ref{eq:Langevin_eq}). In order to calculate 
$\langle \hat{\mathcal{O}} (t_f) \rangle$ one has to sample  a finite number of points $\lbrace \psi_j(t_i) \rbrace_{j=1\ldots N_T}$ from $\sigma_{\mathcal{W}}( \psi^*(t_i), \psi(t_i) )$, let them 
evolve according to the stochastic differential equation (\ref{eq:Langevin_eq}) and then calculate the following expectation value:
\begin{equation}
\begin{array}{rcl}
\langle  \hat{\mathcal{O}}(t_f)  \rangle & \approx & \frac{1}{N_T} \sum\limits^{N_T}_{j=1} \mathcal{O}_{\mathcal{W}}(\psi^*_j(t_f), \psi^{}_j(t_f) ).
\end{array}
\end{equation} 
For large enough $t_f$, a steady state should be reached.\\
One should note, that the strength of the memory kernel in the second term of Eq. (\ref{eq:Langevin_eq}) and the autocorrelation function of the noise depend entirely on the 
properties of the reservoir. Having a reservoir with constant density of states over the entire frequency spectrum,  energy independent couplings $\gamma_k$ and a constant occupation of the modes 
(i.e. $\mathcal{D}(\omega) = \mathcal{D}={\rm const}$, $\gamma_k = \gamma = {\rm const}$, $n(\omega)= n={\rm const}$) the Markovian limit is reached, where the memory kernel vanishes and the stochastic process $\zeta(t)$ 
becomes a Gaussian white noise:
\begin{equation}
\label{eq:Mark_limit}
\begin{array}{rcl}
\partial_t \psi (t) & = & -i\Delta \psi (t) - \Gamma \psi (t)  + \zeta(t) \\[1.5mm]
 \left\langle\zeta(t)\zeta^{\dagger}(t')  \right\rangle &  = & 2 \Gamma(n + \frac{1}{2})\delta(t-t')
\end{array}
\end{equation}
\\
It is important to stress that the same equation is obtained if one starts with the Master equation in Lindblad form for the density matrix $\hat{\sigma}$  of a single quantum dot
\begin{equation}
\label{eq:QMeq}
\begin{array}{rcl}
\partial_t \hat{\sigma} & = & - i [\Delta \hat{a}^{\dagger} \hat{a}] + \hat{\mathcal{L}}\hat{\sigma} \\

\hat{\mathcal{L}}\hat{\sigma} & = & -\Gamma(n+1)[\hat{a}^{\dagger} \hat{a}\hat{\sigma} + \hat{\sigma}\hat{a}^{\dagger} \hat{a} - 2 \hat{a}\hat{\sigma}\hat{a}^{\dagger} ] \\
    &   & -\Gamma n[\hat{a} \hat{a}^{\dagger} \hat{\sigma} + \hat{\sigma}\hat{a}  \hat{a}^{\dagger} - 2 \hat{a}^{\dagger}\hat{\sigma}\hat{a} ],
\end{array}
\end{equation}applies the operator correspondences given in \cite{otago3} in order to map the last expression to a Fokker-Plank equation (FPE) and then use the fact, that the FPE can be rewritten as a Langevin equation. The addition of a dephasing Lindblad operator  $ \mathcal{L} \hat{\sigma} = -\frac{\gamma}{2}[\hat{a}^{\dagger}_{}\hat{a}, [ \hat{a}^{\dagger}_{}\hat{a}, \hat{\sigma} ]] $ to the equation will result only in the appearance of $ \sqrt{\gamma} \psi^*_{}(t) \tilde{\zeta}(t)  $ on the RHS of Eq.  (\ref{eq:Mark_limit}), where $ \tilde{\zeta}(t) $ is a Gaussian white noise $\big( \left\langle \tilde{\zeta}(t) \tilde{\zeta}^{\dagger}_{}(t')  \right\rangle = \delta(t-t') \big) $.\\

The addition of an on-site repulsion term $\frac{U}{2}\hat{a}^{\dagger} \hat{a}^{\dagger} \hat{a} \hat{a}$ 
to the system Hamiltonian reflects in the action $\mathcal{S}''$ by the addition of  
\begin{equation}
 \begin{array}{c}
-U\int d\tau \big[ \big( \psi^{*2}\psi\eta + \eta^* \psi^* \psi^2 \big) +\frac{1}{4}\big( \eta^{*2} \eta \psi + \psi^* \eta^* \eta^2 \big) \big].
 \end{array}
\end{equation}
The $\tau$-dependence is dropped for simplicity here. The terms in the second bracket are neglected to allow for a mapping onto a set of stochastic differential equations. This is the essence of the so called Truncated Wigner Approximation (TWA) \cite{otago1,otago2,otago3}.

\subsection{2.2 Chain of $\mathcal{N}$ quantum dots coupled to two bosonic reservoirs}

The generalisation of the simple example from the previous subsection to the case of an arbitrary number of wells (quantum dots) $\mathcal{N}$ between two reservoirs is straightforward. The Hamiltonian of this system is given by
\begin{equation}
\begin{array}{c}

\hat{H} = 
\sum \limits^{\mathcal{N}}_{j=1} \Delta^{}_{j} \hat{a}^{\dagger}_j \hat{a}^{}_j
+ \sum \limits^{}_{k} \varepsilon^{}_{k}\hat{L}^{\dagger}_{k}\hat{L}^{}_{k}
+ \sum \limits^{}_{k'} \varepsilon^{}_{k'}\hat{R}^{\dagger}_{k'}\hat{R}^{}_{k'}\\
-  \sum \limits^{}_{k}  \gamma^{}_{L,k}\big(   \hat{a}^{\dagger}_1\hat{L}^{}_{k}  +\hat{L}^{\dagger}_{k}\hat{a}^{}_1    \big) 
- \sum \limits^{}_{k'} \gamma^{}_{R,k'}\big(   \hat{a}^{\dagger}_{\mathcal{N}}\hat{R}^{}_{k'} + \hat{R}^{\dagger}_{k'}\hat{a}^{}_{\mathcal{N}}  \big)\\
-\sum \limits^{\mathcal{N}-1}_{j=1} J  \big( \hat{a}^{\dagger}_{j+1}\hat{a}^{}_{j} + \hat{a}^{\dagger}_{j} \hat{a}^{}_{j+1} \big) 
+  \frac{1}{2}\sum \limits^{\mathcal{N} }_{j=1}U^{}_{j} \hat{a}^{\dagger}_{j}\hat{a}^{\dagger}_{j}\hat{a}^{}_{j}\hat{a}^{}_{j}
%
%
\end{array}
\end{equation}
The ladder operators $\hat{L}^{(\dagger)}_{k},\hat{R}^{(\dagger)}_{k},\hat{a}^{(\dagger)}_{}$ are responsible for the annihilation (creation) of an excitation at the left, 
right reservoir and at the chain of quantum dots.
We always set $U_1=0=U_{\mathcal{N}}$ and $\Delta_1 = 0 = \Delta_{ \mathcal{N} }$. \\
The corresponding set of stochastic differential equations that one has to solve is given by
\begin{equation}
\label{eq:Set_of_Stoch_DGL}
\begin{array}{rcl}
\partial_t \psi^{}_1(t) & \! = \! &  - \int^{t}_{t_i}\! d\tau 2 \Gamma_L(t-\tau)\psi^{}_{1}(\tau)
 + iJ\psi^{}_{2}(t) + \zeta^{}_L(t) \\[1.5mm]
\partial_t \psi^{}_{j}(t) & \! = \! &   -i\Delta^{}_j \Psi^{}_j(t) +iJ\big(\psi^{}_{j-1}(t) + \psi^{}_{j+1}(t) \big) \\[1.5mm]
		& & -iU^{}_{j}\psi^2_j(t)\psi^*_{j}(t)   \hspace{23.0mm}(1<j<\mathcal{N}) \\[1.5mm]
\partial_t \psi^{}_{\mathcal{N}}(t) & \! = \! &  - \int^{t}_{t_i} \! d\tau 2 \Gamma_R(t-\tau)\psi^{}_{ \mathcal{N}}(\tau)
 + iJ\psi^{}_{ \mathcal{N }\! - \!  1 \! }(t) + \zeta^{}_{R}(t) 
\end{array}
\end{equation}
where $\Gamma_{L,R},\zeta_{L,R}$ are defined in the same way as in Eq. (\ref{eq:Langevin_eq}) and the subscript $L,R$ refers to the left, right reservoir. 
We assume that initially the lattice chain is empty ($\langle \hat{a}^{\dagger}_i \hat{a}_j \rangle = 0$) and at $t_i=0$ it is instantaneously coupled to the environment, i.e. we take $\gamma^{}_k(t)=\gamma^{}_k \theta(t)$. The Wigner function of the initial state is then
\begin{equation}
\begin{array}{rcl}
\sigma_{\mathcal{W}}(\psi^*, \psi) & = & \prod \limits_j \big( \frac{2}{\pi} e^{-2\psi^*_j \psi^{}_j } \big).
\end{array}
\end{equation}

\section{3 Results for a chain of $\mathcal{N}$ quantum dots coupled to two Markovian reservoirs}
\label{sec:2}

\subsection{3.1 Steady state properties of the system}
\label{subsec:Res_Steady}

We first consider the case $\mathcal{N}=3$. Using the nonequilibrium Green's function approach we get exact results for the noninteracting case and Markovian reservoirs. The mean occupation number $n_j$ $(j=1,2,3)$ of the dots and the steady state current $I$ are given by the following exact solutions, for $\Gamma = \pi \gamma^2 \mathcal{D}$,  $\Delta_2=0$, and $x=J / \Gamma$:
\begin{eqnarray}
n_1 & = & n_L - \frac{n_L-n_R}{2}\frac{x^2}{1+x^2} 	\label{eq:eq_1} ,\\
n_2 & = &  \frac{n_L + n_R}{2} \, ,		\label{eq:eq_2} \\
n_3 & = & n_R - \frac{n_R-n_L}{2 }\frac{x^2}{1+x^2} \, , 	\label{eq:eq_3} \\
I      & = & J\frac{x}{1+x^2}(n_L-n_R)	\, ,		\label{eq:eq_4}
\end{eqnarray}
where $n_{L/R}$ are the occupation numbers of the modes of the left/right reservoir. We should note that the steady state current  remains the same independent of the length of the lattice chain 
as long as $U_j=0=\Delta_j$ $ \forall j $.

In the case of nonzero interparticle interaction in the Markovian limit we approximate the  interaction contribution to the self-energy only by the tadpole diagram (one loop diagram with two external legs, also referred to as Hartree contribution) \cite{Rammer_book}. 
We shall see later that already this approximation yields a number of interesting details, which are consistent with the  TWA predictions.
In the current case the tadpole diagram
renormalises the energy level of the middle quantum dot from $\Delta_2 = 0 $ to $  \Delta_2 =  U_2(1+n_L+n_R) = U_2(1+2n_2) $. At this point it is important to realise that Eq. (\ref{eq:eq_2}) is also valid for $\Delta_2 \neq 0 $, which means that $n_2$ is unchanged in this approximation. The same behaviour of $n_2$ is obtained by the TWA.

\begin{figure}[t]
\centering
\includegraphics[width=0.93\linewidth]{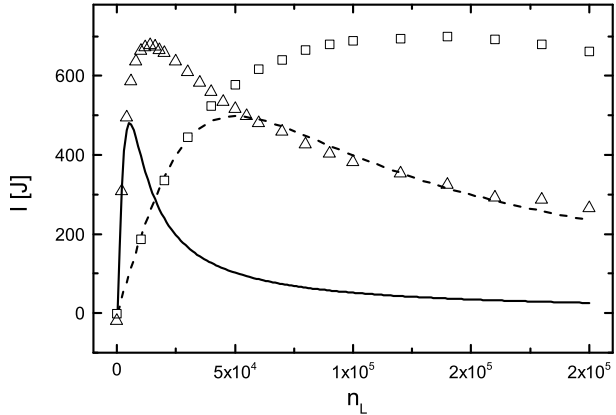}
\caption{\label{fig:Current_ch_Gamma}Steady state current for a chain of three quantum dots coupled to two Markovian reservoirs for nonzero interparticle interaction $U_2/J=10^{-3}$. Truncated 
Wigner approximation (triangles - $\Gamma/J = 5$, squares - $\Gamma/J = 50$) and tadpole approximation (solid line - $\Gamma/J=5$, dashed line - $\Gamma/J=50$). Additional parameters: $\Delta_2 / J=0 $, $n_R=100$. 
The peaks in the tadpole approximation are at $n_L=5 \times 10^{3}$ and $n_L=5 \times 10^{4}$ for $\Gamma/J = 5$ and $\Gamma/J = 50$, respectively. 
The corresponding new values of $\Delta_2/J \rightarrow \Delta_2/J + U_2(1+n_L+n_R)/J \approx  U_2n_L/J $ are $5$ and $50$.}
\end{figure}

\begin{figure}[t]
\centering
\includegraphics[width=1.0\linewidth]{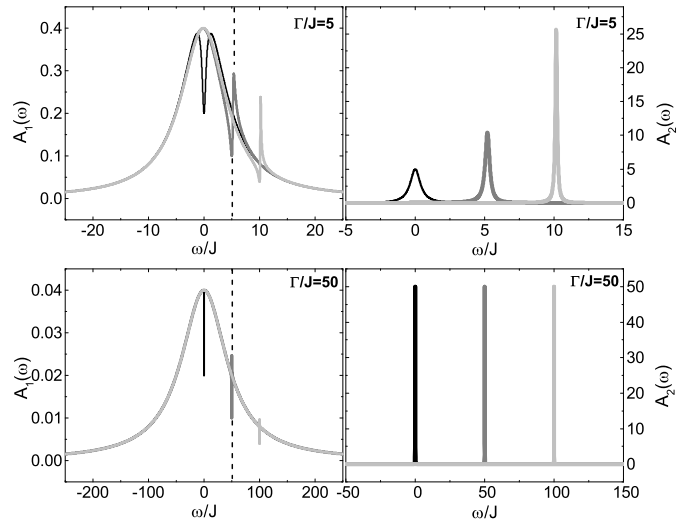}
\caption{\label{fig:Spec_fct_ch_Gamma}
Spectral functions of the first (left panel) and the second (right panel) quantum dot. For $\Gamma/J = 5$ $(50)$ the black, dark grey and grey lines in the upper (lower) two figures denote the spectral functions for $\Delta_2/J = 0$, $5$ and $10$ ($0$, $50$ and $100$) respectively.
The peak of $\mathcal{A}_2(\omega)$  
is at $\omega=\Delta_2$. The dashed vertical line denotes the value of the critical $\Delta_2 \approx U_2n_2$ where the peak in the current in Fig. \ref{fig:Current_ch_Gamma} in the tadpole approximation is reached.}
\end{figure}

From Fig. \ref{fig:Current_ch_Gamma} one sees that the steady state current has qualitatively the same behaviour in the TWA and in the tadpole approximation for not too large $n_L$. The slope of the 
curves and the position of the peaks in the second approximation can be explained with the spectral functions of the three quantum dots 
$\mathcal{A}^{}_j(\omega)$, $(j=1,2,3)$ \cite{Mahan10} that can be obtained from the action $\mathcal{S}'$ of the noninteracting system after the substitution $\Delta_2 \rightarrow \Delta_2 + U_2(1+2n_2)$. In this approximation the spectral functions of the first and third quantum dot are exactly the same since the system is symmetric under the exchange of  $(1,L)\leftrightarrow (3,R)$ indices (except $n_{L,R}$) and the retarded Green's functions of the system do not depend on $n_{L,R}$ in the nonintercating case. In the tadpole approximation this symmetry is not broken since we have only to renormalise $\Delta_2$. For increasing $\Gamma$ $\mathcal{A}_1(\omega), \mathcal{A}_3(\omega)$ become wider and they do not change when varying the energy level $\Delta_2$ of the middle quantum dot, except for the appearance of a small dip and peak at $\omega=\Delta_2$. In the following discussion the latter  effect is not important. On the other hand $\mathcal{A}_2(\omega)$ has only a narrow peak at $\omega=\Delta_2$. \\

Now, we look at the overlap of the spectral functions of the left and the middle quantum dots ($\mathcal{A}_1(\omega), \mathcal{A}_2(\omega)$) (the results for the overlap between 
$\mathcal{A}_2(\omega)$ and $\mathcal{A}_3(\omega)$ are exactly the same). For $\Delta_2 = 0$ and increasing $\Gamma $ the overlap is in the same energy range since the width of $\mathcal{A}_2(\omega)$ 
is almost unchanged in comparison to the width of $\mathcal{A}_1(\omega)$. Only particles in the left dot with energies also accessible in $\mathcal{A}_2( \omega )$ can tunnel to the middle dot. But this number is smaller since, for larger $\Gamma$, $\mathcal{A}_1( \omega )$ spreads over a wider range of energies and the particles at the left dot are distributed over this range. It follows that the current should also decrease. This explains the difference in the slope of the curves plotted in Fig. \ref{fig:Current_ch_Gamma} for small values of $n_L - n_R$. The same behaviour can be seen also in Eq. (\ref{eq:eq_4}) in the relevant parameter regime $x=J/\Gamma \ll 1$.\\

With this simple picture one can also explain the position of the peaks at the curves plotted in Fig. \ref{fig:Current_ch_Gamma}. In the tadpole approximation an increase of the interparticle interaction strength leads to a change of the energy level of the dot ($\Delta_2 \rightarrow \Delta_2 + U_2(1+n_L+n_R) \approx  U_2n_L$). We have to take into account the  competition between two effects. On one hand, an increase of $n_L$ leads to a shift of the peak of $\mathcal{A}_2(\omega)$ to higher values, thus decreasing the overlap between 
$\mathcal{A}_1(\omega)$ and $\mathcal{A}_2(\omega)$, meaning that the relative number of the particles that can tunnel to the middle quantum dot decrease. On the other hand, looking at Eq. (\ref{eq:eq_1}), the total particle number in the first quantum dot increases almost linearly with $n_L$. The position of the peak should be at the point, where the first effect begins to dominate over the second one. From Fig. \ref{fig:Spec_fct_ch_Gamma}, we see that for two different $\Gamma$ this is the value, where $\mathcal{A}_1(\Delta_2 = U_2n_L)$
 is equal to half of its maximum. \\

\begin{figure}[htbp]
 \includegraphics[width=0.94\linewidth]{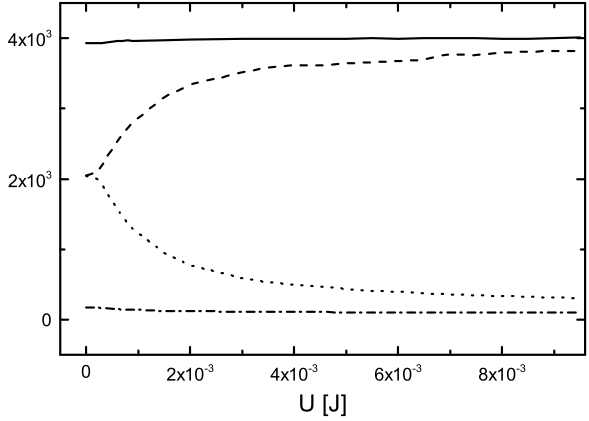}
\caption{\label{fig:Occupation_4Wells_ch_U}
Mean particle occupation of the quantum wells for a chain of four  quantum dots coupled to two Markovian reservoirs, TWA. The black, dashed, dotted and dash-dotted lines denote the mean particle occupation in the first, second, third and fourth quantum dot. We use the parameters: $n_L=4000$, $n_R=100$, $\Gamma /J=5$, $\Delta_1 =\Delta_2=\Delta_3=0$, $U_2=U_3=U$.
}
\end{figure}

Within our approximations and keeping the number of quantum dots $\mathcal{N}=3$, there is no difference in the results for the mean particle occupation  $n_j$ ($1<j<\mathcal{N}$) of the quantum dots in the interacting and noninteracting regime. For $\mathcal{N}\geq 4$ such a difference can be seen as shown in Fig. \ref{fig:Occupation_4Wells_ch_U} for $\mathcal{N}=4$ and $U_2=U_3=U$ after applying the TWA and solving Eq. (\ref{eq:Set_of_Stoch_DGL}). The tadpole approximation cannot describe such a difference in the particle occupation of the middle two dots since it gives the same correction to their energy levels $\Delta_2$ and $\Delta_3$.

We attempted a self-consistent calculation, which leads to the following equations for occupations of the middle two 
quantum dots $(n_2,n_3)=(f_2( \Delta_2,\Delta_3 ),f_3( \Delta_2,\Delta_3 ))$ \cite{Anderson61}:
\begin{equation}
\label{eq:Self_con_eq}
\begin{array}{rcl}
f_2( U(n_2 -1/2) , U(n_3 - 1/2) )  & = & n_2 \\[1.5mm]
f_3( U(n_2 -1/2) , U(n_3 - 1/2) )  & = & n_3 .
\end{array}
\end{equation}
These equations can be solved numerically for a wide set of parameters. In the limit of strong interparticle interactions, we find a better agreement 
of the emerging solutions with the predictions from the TWA for growing $U$.


To explain the results in the strongly interacting limit one has to take into account that each of the Markovian reservoirs forces the occupation in the wells to be equal to the occupation $n_{L/R}$ of the reservoir modes. In the case $\mathcal{N}=4$ and very strong interparticle interactions one should expect that the coupling between the middle two quantum dots is effectively equal to zero in analogy to the self-trapping effect one observes for a Bose-Einstein condensate in a double well potential \cite{Legget01}. One can assume that the first two quantum dots are coupled only to the left reservoir -- and the last two only to the right one. In this case, the occupation of the first two and last two dots is equal to $n_L,n_R$ respectively, which seems to be the case after an extrapolation of the results of both approximations in the limit of large interparticle interaction strengths. 

\subsection{3.2 Transient behaviour of the system}
\label{subsec:Res_Transient}

\begin{figure}[h!tb]
\centering
\includegraphics[width=1.0\linewidth]{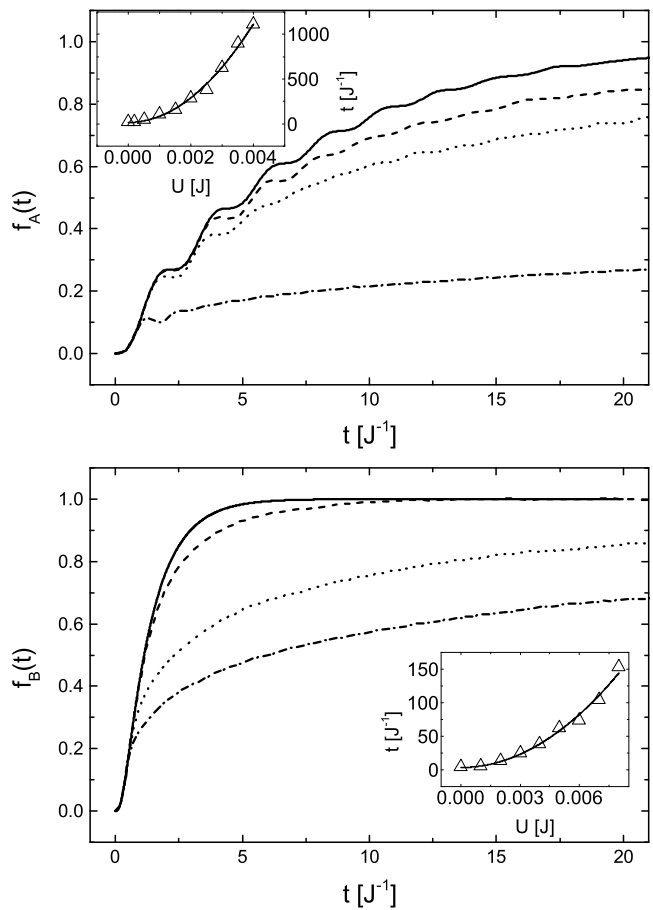}
\caption{
\label{fig:FABB} 
Time evolution of $f_A(t),f_B(t)$ defined in Eq.~(\ref{eq:f_A_and_f_B}) for $U=0$, $\Gamma = \frac{1}{20}2^{3/2}J$ (solid line in the upper panel) and $\Gamma = 5J$ (solid line in the lower panel). The other lines are the results from the TWA obtained after dividing $n_2(t)$ by $(n_L+n_R)/2$. The values of $U_2/J$ are $5 \times 10^{-4},10^{-3},10^{-2}$ ($ 10^{-3},5 \times 10^{-3},10^{-2} $) for the dashed, dotted and dashed-dotted lines in the upper (lower) panel. In the inset one can see the time that $f_A(t)$ or $f_B(t)$ needs to reach $0.95$. The numerical results are fitted with a curve of the form $g(U_2) = a + bU^2_2$.
 }
\end{figure}

In order to find an analytical expression for the behaviour of an empty chain of quantum dots after an instantaneous coupling with two reservoirs one has to calculate the retarded, advanced and lesser Green's function $G^{R,A,<}$ of the system. The case of a single \emph{fermionic} quantum dot coupled to a reservoir is already considered in \cite{Langreth91}, \cite{Schmidt08} in the noninteracting case and in the lowest order self-energy (tadpole) approximation. The generalisation to a chain of quantum dots and two Markovian bosonic reservoirs is straightforward. For $U=0$ the retarded/advanced Green's function is obtained from the solution of the set of equations:
\begin{equation}
\label{eq:Set_for_D_R_A}
\begin{array}{rl}
\big(i\partial^{}_t -  \Delta^{}_l \big)G^{R/A}_{lk}(t,t') & =  \delta_{lk}\delta(t-t') + \\[1.5mm]
    & \sum_j  \int d\tau \Sigma^{R/A}_{lj}(t,\tau)G^{R/A}_{jk}(\tau,t'), \\[1.5mm]
\big( \!\! - \! i\partial^{}_{t'} -  \Delta^{}_k \big)G^{R/A}_{lk}(t,t') & = \delta_{lk}\delta(t-t') + \\[1.5mm]
    & \sum_j  \int d\tau G^{R/A}_{lj}(t,\tau)\Sigma^{R/A}_{jk}(\tau,t').
\end{array}
\end{equation}
The retarded/advanced part of the self-energy has the form
\begin{equation}
\begin{array}{rcl}
\Sigma^{R}_{lk}(t,t') & = & \big(\! - \! i\Gamma \theta(t)( \delta_{l1} \! \delta_{k1} + \delta_{l \mathcal{N}} \!  \delta_{k \mathcal{N} } ) -J\delta_{l,k\pm1}  \big) \delta( \! t \! - \! t' \! ), \\[1.5mm]

\Sigma^{A}_{lk}(t,t') & = & \big(\! + \! i\Gamma \theta(t)( \delta_{l1} \! \delta_{k1} + \delta_{l \mathcal{N}} \!  \delta_{k \mathcal{N} } ) -J\delta_{l,k\pm1}  \big) \delta( \! t \! - \! t' \! ).
\end{array}
\end{equation}
After solving Eq. (\ref{eq:Set_for_D_R_A}) one can obtain $G^{<}(t,t')$ by making use of the fact that the chain of quantum dots is empty at $t=0$:
\begin{equation}
\begin{array}{rcl}
G^{<}(t,t') & = & \int d\tau_1 d\tau_2 G^{R}(t,\tau_1) \Sigma^{<}(\tau_1, \tau_2) G^{A}(\tau_2,t')
\end{array}
\end{equation}
With $G^{R,A,<}(t,t')$ one can obtain all system observables. The calculation of the tadpole approximation of the Green's functions (denoted by $\tilde{G}$) of the chain of quantum dots is obtained via the following equation:
\begin{equation}
\begin{array}{rcl}
\tilde{G}_{lk}(t,t') \! & \! = \! & G_{lk}(t,t') +  \sum \limits_j 2U_j \int_c d\tau n_j(\tau) G_{lj}(t,\tau) G_{jk} (\tau, t').
\end{array}
\end{equation}
The mean occupation number at the $l^{\rm th}$ lattice site is then given by
\begin{equation}
\begin{array}{rcl}
\label{eq:Occupation_Tad}
\tilde{n}_{l}(t) & = & i\tilde{G}^{<}_{ll}(t,t)\\[1.5mm]
		 		 & = & iG^{<}_{ll}(t,t)  + \sum_j 2U_j \int d\tau n^{}_j(\tau) G^R_{lj}(t,\tau)iG^{<}_{jk}(\tau,t) \\[1.5mm]
		 		 &   & \hspace{14.2mm}   + \sum_j 2U_j \int d\tau n^{}_j(\tau) iG^{<}_{lj}(t,\tau)G^R_{jk}(\tau,t).
\end{array}
\end{equation}
The first term is the result from the noninteracting case and the last two are the perturbative corrections from the interaction. In the following we consider the case $\mathcal{N}=3$ and observe only the behaviour of $n_2(t), \tilde{n}_2(t)$.
In the noninteracting case we clearly differ between two regimes in which the observable has the following form:
\begin{equation}
\begin{array}{rcl}
n_2(t) = 0.5(n_L+n_R)f_A(t) \hspace{5.0mm} \Gamma < 2^{3/2}J \\[1.5mm]
n_2(t) = 0.5(n_L+n_R)f_B(t) \hspace{5.0mm} \Gamma > 2^{3/2}J 
\end{array}
\end{equation}
with $f_A(t),f_B(t)$ given by:
\begin{equation}
\label{eq:f_A_and_f_B}
\begin{array}{rcl}
f_A(t) & = & 1 + \frac{e^{-t\Gamma}}{\beta^2}\big( -8J^2 + \Gamma^2 {\rm cos}(t\beta ) - \Gamma \beta {\rm sin}(t \beta )  \big)\\[1.5mm]
f_B(t) & = & 1 + \frac{e^{-t\Gamma}}{\beta^2}\big( 8J^2 - \Gamma^2 {\rm cosh}(t\beta ) -\Gamma \beta {\rm sinh}(t \beta )  \big)\\[1.5mm]
\beta  & = & \sqrt{|8J^2 - \Gamma^2|}.
\end{array}
\end{equation}
In the regime $\Gamma \gg 2^{3/2}J$ (Fig. \ref{fig:FABB}) 
the observable converges exponentially to its steady state, as in the case for the particle occupation of a single quantum dot coupled to a Markovian reservoir. The time scale of this process is proportional to $\Gamma/(4J^2)$. But  in the limit of very small $\Gamma $, one observes a  step-like behaviour of the particle occupation, the length of the steps being equal to $2\pi/\beta$. One can also see that the fastest convergence to a steady state is obtained in the case where $\Gamma \sim J$.\\

The next task is to see if the interparticle interactions at the middle dot can influence this transient behaviour. For the special case of $\mathcal{N}=3$ one can bring Eq. (\ref{eq:Occupation_Tad}) into the more compact form
\begin{equation}
\begin{array}{rcl}
\tilde{n}_2(t) & = & n_2(t) + 4U_2\int d\tau n_2(\tau) \Re \big( G^R_{22}(t,\tau) iG^{<}_{22}(\tau,t) \big).
\end{array}
\end{equation}
The correction to the particle occupation in the middle quantum dot is zero. It follows that not only the steady state but also the transient behaviour of $n_2(t)$ is unchanged by the presence of interactions at least within these approximations. The situation is different if one looks at the numerical solution of Eq. (\ref{eq:Set_of_Stoch_DGL}), where all classical contributions of the interparticle interaction are taken into account. In both parameter regimes $(\Gamma \lessgtr 2^{3/2}J )$ one observes a quadratic dependence of the time needed to reach a steady state from the interparticle interaction in the middle quantum dot (Fig. \ref{fig:FABB}).

\section{4 Conclusions}
\label{concl}

We have studied the transient behaviour and the steady state properties of a chain of quantum dots that is instantaneously coupled to two Markovian reservoirs. For the case of three dots an exact solution in the noninteracting case is shown. We see that the interparticle interaction does not change the mean particle occupation in the middle well in both the TWA and the tadpole approximation. But the time the system needs to reach a stationary state increases quadratically with the  interaction in the TWA. We have also found a qualitative explanation for the behaviour of the steady state current by the use of the spectral properties of the chain of dots. Increasing the number of wells from three to four, additional effects arise from the interparticle interactions. Here the interaction effectively reduces the coupling between the middle two dots such that $n_1 = n_2 = n_L$ and $n_3 = n_4 = n_R$ in the limit of very strong interactions.

In order to access this interesting physics experimentally we envisage the following procedure, which has essentially been partly realized already by the authors of \cite{Anderson12}. One starts with a rather large trap with a Bose-Einstein condensate in perfect equilibrium in it. Then by an instantaneous potential shift one induces a sloshing of the condensate. After that the system shoud be cut into two subsystems, for instance by an impenetrable barrier. In this way one produces two different bosonic reservoirs which contain a large number of particles in excited states. Gradually removing the barrier one can then couple these ``reservoirs'' and hence allow for the transport. The additional structuring of the contact area into several quantum dots can be accomplished in the way similar to that described in \cite{Anderson12} for one well, or by adding a lattice potential along the channel created in \cite{Esslinger12}. We hence expect that such a `bosonic FET' can be manufactured with the state-of-the-art experimental methods.

Needless to say, there is enough room for improvement of our approach.  While an extension of the TWA appears to be highly non-trivial, the inclusion of the higher order self-energies is, in principle, rather straightforward. Since the latter will definitely generate energy-dependent quantities, we expect not only quantitative but also qualitative differences to our predictions to emerge. However, they would only play a significant role for intermediate to strong interactions. 

\section*{Acknowledgements}

We are very grateful to Peter Schlagheck and Martin Bruderer for valuable discussions and for support by the DFG Forschergruppe 760 (Grant No. WI 3426/3-1) and the Heidelberg Center for Quantum Dynamics.

\end{document}